# ENSURING TRANSPORT SECURITY: FEATURES OF LEGAL REGULATION

**Vitaly Khrustalev,**
*Professor, Russian University of Transport,*
*Moscow, Russia*

**Mattia Masolletti,**
*Associate Professor, NUST University*
*Rome, Italy*

**Abstract**

The article analyzes the legal framework regulating the legal provision of transport security in Russia. Special attention is paid to the role of prosecutor's supervision in the field of prevention of crimes in transport.

## 1. Introduction

The relevance of the topic under consideration is due to the specifics and significance of legal relations arising in the transport sector, since their vulnerability always creates a threat (danger) of harming the life and health of a large number of people, a violation of economic stability, as well as public order.

## 2. Main part

The state of transport security in Russia is currently relatively stable. According to the data from the Federal State Statistics Service of Russia ('Rosstat'), 20.1 thousand crimes related to violation of traffic rules and operation of vehicles were registered in 2018, while 7.2 thousand of them caused the death of a person, two or more persons by negligence. In addition, in January-April 2019, 6.3 thousand violations of this type were registered, of which 2 thousand violations resulted in the death of a person, two or more persons by negligence [6]. It should be noted that we are talking only about detected and registered crimes, while it is necessary to realize that there are also latent (hidden) violations in the field of transport security that entail criminal liability. It is also necessary to take into account the degree and number of administrative offenses in the field of transport security, for which administrative responsibility is established. Questions about the state of transport security in Russia also come from the annual report of the Prosecutor General of the Russian Federation at a meeting of the Federation Council of the Federal Assembly. Thus, it follows from this report that '… over the past year, the work has continued to ensure the rule of law in the transport sector, especially in matters of passenger safety. Attention was drawn to the strengthening of the inspection regime at thousands of facilities, the training and certification of more than 6 thousand employees of security services, the suspension of the operation of aircraft that do not meet the requirements of airworthiness. The tasks for the current year remain flight safety, control over the implementation of programs for the development of the transport industry, respect for the rights of passengers, suppression of the facts of importation into the country of products for which bans and restrictions are established' [5].

The following regulatory legal acts form the legal basis for regulating transport security at the federal level:

1) Federal Law No. 16-FZ 'On Transport Security' (adopted on February 9, 2007). This law is fundamental in the issue of transport security. It defines the provisions related to the information support of transport security, the rights and obligations of transport infrastructure and security entities, the procedure for the inspection of transport, individuals, baggage and cargo to ensure transport security, as well as issues of training, certification and accreditation of transport security forces and their units. This law also establishes such important legal terms as: transport security, an act of unlawful interference and others.

2) 'The Air Code of the Russian Federation' N 60-FZ (adopted on March 19, 1997). It regulates relations arising in connection with activities in the field of aviation, the performance of aircraft flights, the establishment of the procedure for ensuring aviation security, licensing and certification in the field of aviation, and others.

3) 'The Code of Inland Water Transport of the Russian Federation' N 24-FZ (adopted on March 7, 2001). It regulates relations in the field of navigation on the inland waterways of the Russian Federation, as well as establishes the procedure for the carriage of passengers, cargo, baggage, rescue of ships and other property, safety of navigation, and actions in case of an accident.

4) 'The Commercial Navigation Code of the Russian Federation' N 81-FZ (adopted on April 30, 1999). It regulates relations related to the implementation of commercial navigation and the provision of sea transportation, as well as defines provisions related to accidents and the implementation of state port control, and other.

5) Federal Law No. 17-FZ 'On Railway Transport in the Russian Federation' (adopted on

January 10, 2003). It regulates relations on the use of public railway transport, the basics of state regulation in the field of non-public railway transport, establishes the basics of safety in railway transport, ensures the protection of goods, railway transport facilities and other.

6) Federal Law No. 259-FZ 'Charter of motor transport and urban ground electric transport' (adopted on November 8, 2007). This law defines the general conditions for the transportation of passengers and their luggage, cargo, as well as the general conditions for the provision of such services at transport infrastructure facilities and others.

7) Other Federal Laws.

The secondary legislation in the field of transport security include: decrees of the President of the Russian Federation, resolutions of the Government of the Russian Federation, orders of the Ministry of Transport of the Russian Federation. The most significant of them are:

1) Decree of the President of the Russian Federation No. 403 'On the creation of an integrated system for ensuring public safety in transport' (adopted on March 31, 2010);

2) Order of the Government of the Russian Federation N 1285-r 'On approval of the Comprehensive Program for ensuring public safety in transport' (adopted on July 30, 2010; in edition of December 11, 2013);

3) Decree of the Government of the Russian Federation N 1090 'On Traffic Regulation' (adopted on October 23, 1993; in edition of February 13, 2018);

4) Decree of the Government of the Russian Federation N 880 'On approval of the Regulations on Federal state control (supervision) in the field of transport security' (adopted on October 04, 2013; in edition of February 17, 2018);

5) Decree of the Government of the Russian Federation N 495 'On approval of requirements for ensuring transport security, including requirements for anti-terrorist protection of objects (territories) that take into account security levels for various categories of transport infrastructure objects and railway transport vehicles' (adopted on April 26, 2017);

6) Decree of the Government of the Russian Federation No. 410 'On approval of requirements for ensuring transport security, including requirements for anti-terrorist protection of objects (territories) that take into account security levels for various categories of subways' (adopted on April 5, 2017);

7) Decree of the Government of the Russian Federation No. 678 'On requirements for ensuring transport security, including requirements for anti-terrorist protection of objects (territories) that take into account security levels for various categories of transport infrastructure objects and sea and river transport vehicles' (adopted on July 16, 2016);

8) Decree of the Government of the Russian Federation No. 1208 'On approval of requirements for compliance with transport security for individuals following or staying at transport infrastructure facilities or vehicles, by type of transport' (adopted on November 15, 2014; in edition of October 3, 2015);

9) And others. It is necessary to separately highlight such subordinate normative acts as the 'Concept of Public Security in the Russian Federation" approved by the President of the Russian Federation on November 14, 2013 (President's Decree N 2685) and the Decree of the President of the Russian Federation N 683 'On the National Security Strategy of the Russian Federation' (adopted on December 31, 2015; the latest edition – July 2, 2021). These acts indicate a deterioration in the technical condition of transport infrastructure facilities, vehicles, which in turn entails the risk and possibility of emergencies and crimes. In addition, there is a mention that one of the tasks to ensure public safety is to improve road safety, reduce the number of road accidents that cause harm to the life and health of citizens, reduce the severity of their consequences.

It is also indicated that it is necessary to prevent road accidents, crimes and other offenses committed by negligence in everyday life, on transport. State control is a necessary element of public administration of the company. It is a kind of 'measure' of the effectiveness of state influence on public relations in certain areas of activity. It is in the process of control (supervision) that it is possible to identify how effective public administration is. All countries of the world form systems of state control operating in various areas of society's life. At the same time, as the analysis of international practice shows, these systems are characterized by a set of the following features: strict regulation of the procedure for carrying out control activities of the state; independence of state control bodies; high certainty of the object and subject of control; uniform legal regulation of control mechanisms; openness and transparency of methods of control and supervisory activities of state authorities [3; 8-10].

The President of the Russian Federation, Vladimir Putin, when speaking at a meeting of the Board of the Prosecutor General's Office on March 14, 2017, has mentioned that citizens judge the ability of the state itself to protect their interests, to defend truth and justice precisely by the effectiveness of the work of prosecutors [7]. Nowadays, prosecutor's supervision is an integral part of the activities of law enforcement agencies to ensure law and order, legality and elimination of various offenses, especially in the transport sector. The Prosecutor's Office of the Russian Federation is a unified federal centralized system of bodies that, on behalf of the Russian Federation, oversee compliance with the Constitution of the Russian Federation and the execution of laws in force on the territory of the Russian Federation, as well as other functions established by federal laws in order to ensure the rule of law, unity and strengthening of the rule of law, protection of human and civil rights and freedoms, as well as the interests of society and the state protected by law [1]. The Transport Prosecutor's Office is a state law enforcement agency that oversees compliance with the law and investigates crimes committed on railway, water, and air transport. It is a specialized prosecutor's office of the Russian Federation.

According to their structure, transport prosecutor's offices are of three groups:

1) District transport Prosecutor's offices.

2) Regional transport prosecutor's offices on the rights of regional ones.

3) The Department for Supervision of the Implementation of Laws on Transport and in the Customs Sphere.

The main powers of transport prosecutor's offices equated to the prosecutor's offices of the subjects of the Russian Federation include:

1) Supervision of the implementation of the Constitution of the Russian Federation, verification of compliance with the law by the transport management bodies.

2) One of the main areas of activity is criminal prosecution, which is carried out in three categories of cases, namely: crimes related to transport organizations and related directly to transport; crimes that are committed due to negligence or non-performance of their official duties by an employee of a transport organization; crimes that are related to violation of traffic rules on transport. The transport prosecutor participates both in pre-trial proceedings and directly in court, presenting the charge to the defendant.

3) The transport prosecutor also participates in the proceedings of civil and arbitration cases.

4) The Transport Prosecutor participates in the proceedings of administrative cases and cases related to the customs sphere.

5) The Transport Prosecutor should exchange information with the prosecutors of the constituent entities of the Federation, interact with them and jointly coordinate activities to combat offenses.

6) Interact with the media.

7) Consider appeals from citizens in the field of transport in case of violation of the law.

8) Conducting an inspection of the customs authorities. Specialized prosecutor's offices, including transport, are available in every federal district: in the North-Western, Volga, Central, Ural, Siberian, Far Eastern, North Caucasus, and Southern Federal Districts.

The control and supervisory measures are carried out in relation to transport infrastructure entities, on the basis of an annual inspection plan, as a rule, by complex commissions, including representatives of the Ministry of Internal Affairs and the Federal Security Service of Russia, transport prosecutor's offices. 'In specialized prosecutor's offices, the prosecutor must be not only a good lawyer, but also a technician, an economist, and so on', once noted the former Prosecutor General of the Russian Federation Y. Chaika in his interview to 'Rossiyskaya Gazeta' (the 'Russian Newspaper') [4]. The Prosecutor General draws attention to the fact that the transport prosecutor must have a deeper knowledge of the subject than the territorial one [4]. Indeed, the Transport Prosecutor's Office is consistently working to protect and restore the rights of citizens, the interests of society and the state. As an example, we can take the Southern Transport Prosecutor's Office, whose staff recently celebrated the decade of education.

On April 27, 2007 the Prosecutor General had issued an order 'On the formation of the Southern Transport Prosecutor's Office" [2]. For 14 years, its employees have identified about 400 thousand violations of laws, in order to eliminate them, more than 50 thousand acts of prosecutor's response were introduced, 20 thousand officials and legal entities were brought to administrative responsibility. According to the materials of inspections sent to the preliminary investigation bodies, more than a thousand criminal cases have been initiated, for which the courts have entered guilty verdicts. Prosecutors have taken measures aimed at suppressing the activities of organizations and persons providing services of inadequate quality and posing a threat to the life and health of citizens, including passenger transportation and vehicle maintenance. Interdepartmental working groups operate on an ongoing basis, the results of the investigation of each accident during the operation of transport are studied by prosecutors, the actions of officials of enterprises and transport organizations, state and regulatory bodies authorized in this area are evaluated.

### 3. Conclusion

Thus, analyzing the above, we can conclude that transport security and its provision is one of the important activities of the state, including the prosecutor's office, since transport security is a part of national security, therefore, its destabilization threatens the interests of not only the state, but also society as a whole.

# References

[1] Federal Law No. 2202-1 'On the Prosecutor's Office of the Russian Federation' (adopted on January 17, 1992; amended on 27.12.2018).

[2] Order of the Prosecutor General's Office of the Russian Federation No. 70 'On the formation of the Southern Transport Prosecutor's Office' (adopted on April 27, 2007).

[3] Administrative reform in Russia (2006). Scientific and practical guide. Edited by S. Naryshkin, T. Ya. Khabrieva.

[4] Interview of the Prosecutor General of the Russian Federation Yuri Chaika to Rossiyskaya Gazet. URL: https://genproc.gov.ru/special/smi/news/news-58850 /

[5] Report of the Prosecutor General of the Russian Federation at the meeting of the Federation Council of the Federal Assembly of the Russian Federation (dated April 18, 2018). URL: http://genproc.gov.ru/genprokuror/appearances/document-1367434

[6] Federal State Statistics Service (2021). URL: http://www.gks.ru/wps/wcm/connect/rosstat_main/rosstat/ru/statistics/population/infraction

[7] Meeting of the Board of the Prosecutor General's Office of Russia. URL: http://www.kremlin.ru/events/president/news/54035

[8] Pravkin, S., Smirnova, V., Bogdanova, Y., Belozerova, I., Selezneva, N., Nasonkin, V. (2019) Effective legal management of investments – the basis of sustainable economic development. Revista Turismo Estudos & Práticas (RTEP), Mossoró/RN, Caderno Suplementar 01. pp.1-9.

[9] Pravkin, S., Smirnova, V., Shagieva, R., Arkhipov, I., Erofeeva, D. (2020) Meso-level investment management in transport projects with the application of the public-private partnership mechanism // E3S Web of Conferences 208, 06019. DOI: doi.org/10.1051/e3sconf/202020806019.

[10] Taradonov, S. et al. (2019) The mechanism of legal regulation in the conditions of globalization and formation of information environment. Regional aspect // Journal of Environmental Management and Tourism. – Vol. 10. № 7 (39). pp. 1517-1521.